\documentclass[aps,prl,superscriptaddress,twocolumn,twoside,a4paper,floatfix]{revtex4}
\usepackage{hyperref,graphicx,amsmath,amssymb,mathpazo,color}
\usepackage[normalem]{ulem}

\usepackage{amsfonts,epsfig}
\usepackage{latexsym}
\usepackage{epsfig}
\usepackage{amsmath}
\usepackage{amssymb}
\usepackage{amsfonts}
\usepackage{amsthm}
\usepackage{mathrsfs}
\usepackage{natbib}
\usepackage{color,verbatim,graphics}
\usepackage{psfrag}
\DeclareMathAlphabet{\mathrsfs}{U}{rsfs}{m}{n}
\DeclareMathAlphabet{\mathpzc}{OT1}{pzc}{m}{it}
\DeclareMathAlphabet{\matheus}{U}{eus}{m}{n}
\DeclareMathAlphabet{\mathbbold}{U}{bbold}{m}{n}

\newcommand{\ba}{\begin{eqnarray}}
\newcommand{\be}{\begin{equation}}
\newcommand{\ee}{\end{equation}}

\newcommand{\ea}{\end{eqnarray}}
\newcommand{\ban}{\begin{eqnarray*}}
\newcommand{\ean}{\end{eqnarray*}}

\newcommand{\ket}[1]{|#1\rangle}

\begin{document}

\title{Bell nonlocality and Bayesian game theory}

\author{Nicolas Brunner}
\affiliation{H.H. Wills Physics Laboratory, University of Bristol, Bristol, BS8 1TL, United Kingdom}
\affiliation{D\'epartement de Physique Th\'eorique, Universit\'e de Gen\`eve, 1211 Gen\`eve, Switzerland}
\author{Noah Linden}
\affiliation{School of Mathematics, University of Bristol, Bristol BS8 1TW, United Kingdom}

\begin{abstract}
We discuss a connection between Bell nonlocality and Bayesian games.  
This link offers interesting perspectives for Bayesian games, namely to allow the players to receive advice in the form of nonlocal correlations, for instance using entangled quantum particles or more general no-signaling boxes. The possibility of having such 'nonlocal advice' will lead to novel joint strategies, impossible to achieve in the classical setting. This implies that quantum resources, or more general no-signaling resources, offer a genuine advantage over classical ones.
Moreover, some of these strategies can represent equilibrium points, leading to the notion of quantum/no-signaling Nash equilibrium.
Finally we describe new types of question in the study of nonlocality, namely the consideration of non-local advantage when there is a set of Bell expressions.
\end{abstract}

\maketitle

On several occasions in the history of science, different areas of research, sharing a priori nothing in common and sometimes belonging to completely different fields of science, were shown to be closely related. In certain cases, these links turned out to be spectacular and tremendously fruitful, such as the connection between differential geometry and relativity.
In the present paper, we discuss such a link, albeit a much more modest one, between Bell nonlocality and the theory of Bayesian games---also referred to as games with incomplete information.  

Nonlocality is arguably among the most dramatic and counter-intuitive features of quantum mechanics. 
In a nutshell, quantum theory is at odds with the principle of locality, which states that an object is influenced directly only by its immediate surroundings, and not by remotely located objects. Two remote observers sharing a pair of entangled quantum particles, can establish correlations which evade any possible explanation in classical physics. On the one hand, a signal is excluded, as it would have to travel faster than light. On the other, the correlated behaviour is not the result of a pre-established strategy, as demonstrated by Bell in 1964 \cite{bell2}. 
Notably, this phenomenon of quantum nonlocality, confirmed experimentally \cite{aspect} via the violation of so-called Bell inequalities, turns out to be useful in practice, in particular for information processing \cite{ekert,cc}. More recently a theory of generalized nonlocal correlations has been developed \cite{pr}, which has direct impact on fundamental questions in the foundations of quantum mechanics \cite{popescu}.

In a completely different area, but only 3 years after Bell's ground-breaking discovery, Harsanyi \cite{harsanyi} developed a framework for games with incomplete information, that is, games in which players have 
only partial information about the setting in which the game is played. For instance, each player may have some private information, such as his payoff, unknown to other players. Harsanyi's discovery marked the start of Bayesian game theory, which now plays a prominent role in game theory and in economics, used in particular to model auctions.

Here we discuss a strong connection between Bayesian games and Bell nonlocality. Specifically, the normal form of a Bayesian game can be reformulated as a Bell (inequality) test scenario. Central to our study will be the possibility for the players to use a common piece of advice (originating e.g. from an advisor), allowing for correlated strategies. The kind of physical resources available to the advisor limits the possible strategies of the players. Notably, players sharing nonlocal resources, such as entangled quantum particles, can outperform players having access to the most general classical resources. This advantage of nonlocal resources occurs for instance when the payoff function of the players corresponds to a Bell inequality, as first discussed by Cheon and Iqbal \cite{cheon}, and further developed in Refs \cite{more}. However, we shall see that there exist more general situations, in which none of the payoffs functions corresponds to a Bell inequality, where nonlocal resources (in particular entanglement) provide nevertheless an advantage over any classical strategy.
Notably some of these nonlocal strategies represent equilibrium points, termed quantum Nash equilibria or non-signaling Nash equilibria. To illustrate these ideas, we discuss several simple examples.

Finally, we emphasize that, for the class of games discussed here (i.e. Bayesian games), quantum mechanics provides a clear and indisputable advantage over classical resources, in the most general sense. This is in contrast with some previous approaches to quantum games \cite{eisert}, based on non-Bayesian games (or complete games, such as Prisoner's dilemma), for which a quantum advantage is achieved only under specific restrictions, the relevance of which has been much debated \cite{comment}. 


\section{From Bayesian games to Bell inequalities}

Let us start with the normal form representation of a game. At this stage, one needs to specify the number of players, the set of possible strategies for each player, and the payoff function for each player. To model Bayesian games, Harsanyi proposed to introduce Nature as an additional player to the game. 
In particular, Nature assigns to each player a \emph{type}, chosen from a given set. 
The type of each player is generally unknown to other players, and determines, for instance, his payoff function. 
The set of possible types for each player is thus also part of the definition of a Bayesian game. More formally, the normal form representation of a Bayesian game is given by the following ingredients \cite{osborne}:

\begin{enumerate}
\item The number of players $N$.
\item A set of states of nature $\Omega$, with a prior $\mu(\Omega)$.
\item For each player $i$, a set of actions $\mathcal{A}_i$.
\item For each player $i$, a set of types $\mathcal{X}_i$.
\item For each player $i$, a mapping $\tau_i: \Omega \rightarrow \mathcal{X}_i$.
\item For each player $i$, a payoff function $f_i: \Omega  \times \mathcal{A}_1 \times ... \times \mathcal{A}_N \rightarrow  \mathbb{R}$, determining the score of the player for any possible combination of types and actions.
\end{enumerate}

In the following, we shall focus on the case of two players, for simplicity. The variable $A_1=0,1,...k_1$ denotes the possible actions of
party 1, and $X_1=0,1,...m_1$ denotes the possible types of party 1 etc.
We will also consider that the possible states of Nature are simply the combination of all possible types (for all players), that is $\Omega = (\mathcal{X}_1, \mathcal{X}_2)$.
In order to play the game, each player should decide on a particular strategy to follow. A pure strategy then consists in associating an action for every possible type, i.e. a mapping $s_i: \mathcal{X}_i \rightarrow \mathcal{A}_i$. More generally, players may use a probabilistic strategy, hence it is convenient to define a probability of an action given a type, i.e. $P(A_i|X_i)$. 
An important feature of the game is then the average payoff function, or the average score, for each player. For player $i$ this is given by

\ba \label{payoff} F_i = \sum \mu(X_1,X_2) P(A_1,A_2|X_1,X_2) f_i(X_1,X_2, A_1 , A_2) \ea
where the sum goes over all variables $X_1,X_2,A_1,A_2$.
Note that if the players use pure or independent strategies, then $P(A_1,A_2|X_1,X_2)= P(A_1|X_1)P(A_2|X_2)$. However, in certain cases 
the players may adapt their strategy depending on a piece of advice. The latter is delivered to all players by an advisor. This opens the possibility for the players to adopt correlated strategies, which can outperform independent strategies. 
There are various forms that advice can take. For example in the case of correlated classical advice, the advice
is represented by a classical variable, $\lambda$, with prior $\rho(\lambda)$. Each player can then choose a strategy depending on his type and on $\lambda$. In general for classical correlated strategies, we have that 
\ba P(A_1,A_2|X_1,X_2) &=& \sum_{\lambda} \rho(\lambda) P(A_1|X_1,\lambda)P(A_2|X_2,\lambda) \nonumber \\
& \neq & P(A_1|X_1)P(A_2|X_2) \ea
An important point in what follows (and not just in
the case of classical advice) is that the advice must be independent of the state of Nature, that is, the choice of types is unknown to the advisor. This enforces the following condition
\ba \label{NS} P(A_1|X_1,X_2) = \sum_{A_2}  P(A_1,A_2|X_1,X_2) = P(A_1|X_1) \ea
which states that the marginal of player 1 does not depend on the type of player 2; a similar condition holds for the marginal of player 2. In the context of Bell nonlocality (as we shall see below), the above condition plays a prominent role. It is referred to as the 'no-signaling' condition, which imposes that the correlations $P(A_1,A_2|X_1,X_2)$ do not allow for player 2 to signal instantaneously to player 1, and thus respect causality and are not in conflict with relativity. We note that in the context of games, the situation in which the advice can depend on the types has been considered \cite{forges}; physically this situation is however not so relevant for us, as it involves signaling.


To analyze games it is often useful to define the set of all possible pairs of payoff functions $\{F_1,F_2\}$, considering all possible strategies. It is convenient to represent geometrically the space of possible payoff functions \cite{osborne}, i.e. here the set of points in $\mathbb{R}^2$ with coordinates $(F_1,F_2)$. In case the players share classical advice (and the number of possible strategies is finite) this space is a convex polytope. The space can then be conveniently characterized with a finite set of linear inequalities of the form
\ba \label{facets} \sum_{j=1}^{N=2} \beta_j F_j \leq \beta_0 \ea
where $\beta_j$ are real numbers. These inequalities define the facets of this polytope.

A notion of particular importance in game theory is that of a Nash equilibrium. In the case the game features an advisor, there is a more refined notion of correlated equilibrium \cite{aumann}. Players achieve a correlated equilibrium for a given set of strategies if each player has no incentive to change strategy, that is, his average payoff will not increase by choosing any other possible strategy (keeping the other player's strategy fixed).

We shall see now that the above scenario is closely related to that of a Bell test. For simplicity, we will focus on a Bell test with two parties, although more parties can be considered. We consider two separated observers, Alice and Bob, sharing a physical resource distributed by a central source (see Fig.~1). 
Each observer receives a question to which she/he is asked to give an answer. In more physical terms, these questions should be understood as measurement settings, and the corresponding answers as measurement outcomes. Importantly each observers knows only his own question, and does not know which question the other observers receive.
To make the analogy with a Bayesian game, the questions here correspond to the type of each player, while the answers correspond to the actions. Hence we will denote by $X_1$ and $X_2$ the questions of Alice and Bob, respectively, and by $A_1$ and $A_2$ the corresponding answers.
After repeating the above operation a large number of times, the statistics of the experiment can be computed, resulting in the joint probability distribution 
\ba \label{joint} P(A_1,A_2|X_1,X_2) \ea
which represents the probability of observing a pair of answers $A_1,A_2$, given a pair of questions $X_1,X_2$.

 \begin{figure}[t!]
 \includegraphics[width=0.9\columnwidth]{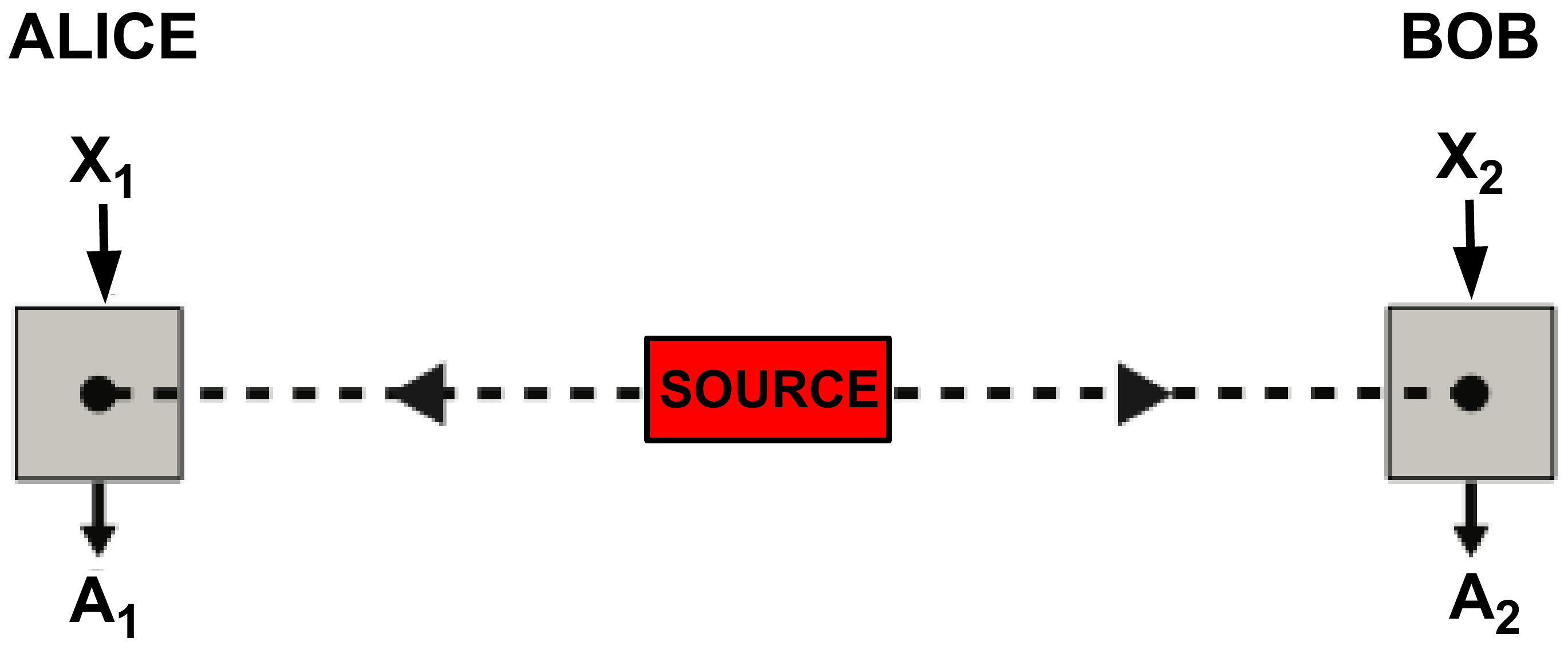}
  \caption{Bell inequality test scenario.}
\label{fig}
\end{figure}

In a Bell test, the goal is loosely speaking to capture the strength and the nature of the correlations observed in the experiment. In general this may depend on the kind of physical resource distributed by the source to the observers. 
A case of particular importance is that of a classical source (i.e. a source of classical particles). In particular, the particles can be thought of as carrying information about a common strategy, which will eventually lead to a correlated behaviour in the experiment. 
The statistics of any experiment involving a classical source can be written as
\ba \label{local} P(A_1,A_2|X_1,X_2) = \int d\lambda \rho(\lambda) P(A_1|X_1,\lambda) P(A_2|X_2,\lambda) \ea
where the variable $\lambda$ (distributed according to the prior $\rho(\lambda)$, with $\int d\lambda \rho(\lambda)=1 $) represents the common strategy, that is, the information distributed from the source to all the observers. From the point of view of games, the variable $\lambda$ represents the advice. Thus the source models the advisor. Importantly, all possible strategies for players receiving classical advice (i.e. from a classical advisor) are of this form.

In his 1964 ground-breaking work, Bell discovered that, the correlations obtained in any experiment involving a classical source are constrained. More formally, any statistics of the form \eqref{local} satisfies a set of inequalities, now known as Bell inequalities \cite{bell2}. Generally, a Bell inequality is based on a linear expression of the joint probabilities \eqref{joint}, of the form
\ba \label{bell} S = \sum_{X_1,X_2,A_1,A_2} \alpha_{X_1,X_2,A_1,A_2} P({A_1,A_2|X_1,X_2}) \ea
where $\alpha_{X_1,X_2,A_1,A_2}$ are real numbers.
The maximum of $S$ over all possible strategies of the form \eqref{local} is called the local bound of the inequality $L$. Putting all together, a Bell inequality then reads
\ba \label{BI} S \leq L. \ea

Now we come to an important point. The average payoff function (for a given player $i$) is essentially a Bell expression. Indeed, Eqs \eqref{payoff} and \eqref{bell} are exactly of the same form, with 
\ba \mu(X_1,X_2) f_i(X_1,X_2,A_1,A_2) = \alpha_{X_1,X_2,A_1,A_2} \ea
Hence to any average payoff function $F_i$ can be associated a Bell expression $S_i$. Moreover, in the presence of a classical advisor, the following condition must hold: $F_i\leq L$, where $L$ is the local bound of the Bell expression associated to $F_i$.

More generally, note that the above reasoning also applies to any linear combination of the payoff functions. In particular, the facets of the space of payoff functions, of the form \eqref{facets}, can also be associated to a Bell expression $S$. Thus, in the case of a classical advisor, the condition \eqref{facets}, can be seen as a Bell inequality, with Bell expression $S=\sum \beta_j F_j $ and local bound $L=\beta_0$.

To summarize, the payoff function of a Bayesian game is basically a Bell expression, and is hence limited by Bell's inequality for any strategy involving a classical advisor. More generally, this applies to linear combinations of payoff functions, such as those corresponding to the facets of the space of payoff functions. Next we shall move to quantum mechanics, for which the situation turns out to be dramatically different!

Remarkably, in experiments involving a source of quantum particles, Bell's inequality \eqref{BI} can be violated. This means that there exist quantum experiments, the statistics of which cannot be written in the form \eqref{local}. This is quantum nonlocality, a phenomenon repeatedly observed experimentally which has many applications in quantum information processing.

A crucial feature of quantum correlations is that they satisfy the no-signaling principle, represented by conditions of the form \eqref{NS}. This is indeed fundamental, as it ensures that quantum mechanics is compatible with relativity. It turns out however that there exist correlations which are stronger than those allowed in quantum mechanics, which nevertheless satisfy the no-signaling principle. Such correlations, discovered by Popescu and Rohrlich \cite{pr}, are often referred to as 'super-quantum correlations' or 'nonlocal boxes'.

For Bayesian games, the possibility of having access to nonlocal correlations, for instance using entanglement, has important implications. First let us imagine that the players can share quantum advice, that is, the advisor is able to produce entangled particles and to send them to the players, who then perform local measurements on their particles. Since the statistics of such measurements can in general not be reproduced by any classical local model, the players now have access to strategies which would be impossible in the case of a classical advisor. Thus, players sharing quantum advice can outperform \emph{any} classical players. More formally, this means that the space of payoff functions for players sharing quantum advice can become larger then the space of payoff functions for classical players. In case one the average payoff function of one (or more) player corresponds to a Bell inequality, then quantum resources give an advantage to the players. Interestingly however, even in the case none of the payoff functions corresponds to a Bell inequality (i.e the highest possible payoff can be reached classically), it is still possible in certain cases to obtain a quantum advantage.

Going beyond quantum mechanics, it is relevant to consider general nonlocal resources in the context of Bayesian games. In general, this allows for novel strategies, which can outperform both classical and quantum strategies. Hence the space of payoffs achievable with no-signaling strategies is in general larger than in the case of quantum strategies.

Finally, allowing for quantum or super-quantum strategies also provides novel correlated equilibrium points to the game. Such points are referred to as quantum Nash equilibria and no-signaling Nash equilibria.

Below we will illustrate these ideas by discussing a few simple examples of Bayesian games featuring a 'quantum advantage' and a 'no-signaling advantage'.


\section{Examples}

{\bf \emph{Example 1.}} We first consider a simple game between two players, characterized as follows. For each player there are only two possible types, $X_1=0,1$ for the player 1 (from now called Alice), and $X_2=0,1$ for the player 2 (from now called Bob). The set of possible actions is also composed of two elements only: actions $A_1=0,1$ for Alice, and $A_2=0,1$ for Bob.
There are thus four possible states of Nature, and we will consider them equally likely: $\mu(X_1,X_2)=1/4$ for $X_1,X_2=0,1$.

Next we define the payoff function of Alice to be given by 
\ba f_1(X_1,X_2,A_1,A_2) = 
\begin{cases}
+4 & \text{if $ A_1 \oplus A_2  = X_1 X_2$} \\
-4 & \text{otherwise}
\end{cases}
\ea
where $\oplus$ designates addition modulo 2.
Thus the average payoff function of Alice is given by

\ba \label{chsh} F_1 &=& E(X_1=X_2=0) +  E(X_1=0,X_2=1) \nonumber \\ & +&  E(X_1=1,X_2=0) -  E(X_1=X_2=1) \ea
where we have defined the correlation function 
\ba \nonumber E(X_1,X_2)= P(A_1=A_2 |X_1,X_2)-P(A_1 \neq A_2 |X_1,X_2) \ea

We will consider the game to be symmetric hence the payoff function of Bob is the same as that of Alice, i.e. $F_1=F_2$.

Now it turns out that the function \eqref{chsh} is very well known in quantum mechanics. It is the basis of the simplest Bell inequality, derived in 1969 by Clauser-Horne-Shimony-Holt (CHSH) \cite{chsh}. The CHSH Bell inequality reads $F_1 \leq 2$. In quantum mechanics, by performing judicious measurements on a singlet state, of the form $\ket{\psi_-}= (\ket{0}_A \ket{1}_B-\ket{1}_A \ket{0}_B)/\sqrt{2}$, it is possible to obtain the following set of correlation functions:
\ba \label{Qopt} E_{\psi_-}(X_1,X_2) =  (-1)^{X_1X_2}\frac{1}{\sqrt{2}} \ea
This leads to $F_1=2\sqrt{2} >2$, hence violating the CHSH inequality. 
Therefore, the space of possible payoffs in the case of quantum advice is clearly larger than in the classical case, since there exist individual values the payoff function which are not attainable classically.

Moreover, the quantum setting provides here a new equilibrium point. Indeed it turns out that the value of $F_1=F_2=2\sqrt{2}$ is in fact the maximum that is achievable in quantum mechanics \cite{tsirelson}. 
Therefore the point $F_1=F_2=2\sqrt{2}$ represents a quantum correlated equilibrium point, as it is impossible for Alice or Bob to obtain a larger payoff by adopting any other strategy.

Next let us consider super-quantum correlations. It turns out that such correlations can give rise to maximal violation of the CHSH inequality, reaching CHSH=4. Thus, players sharing such 
super-quantum correlations can outperform quantum players in the above game, and reach $F_1=F_2=4$, achieving the highest possible average payoff. In particular this is achieved using no-signaling correlations known as the Popescu-Rohrlich (PR) box, characterized by $E_{\text{PR}}(X_1,X_2) =  (-1)^{X_1X_2}$. Indeed the point $F_1=F_2=4$ is a no-signaling Nash equilibrium, since no higher payoffs are possible.

 \begin{figure}[t!]
 \includegraphics[width=0.9\columnwidth]{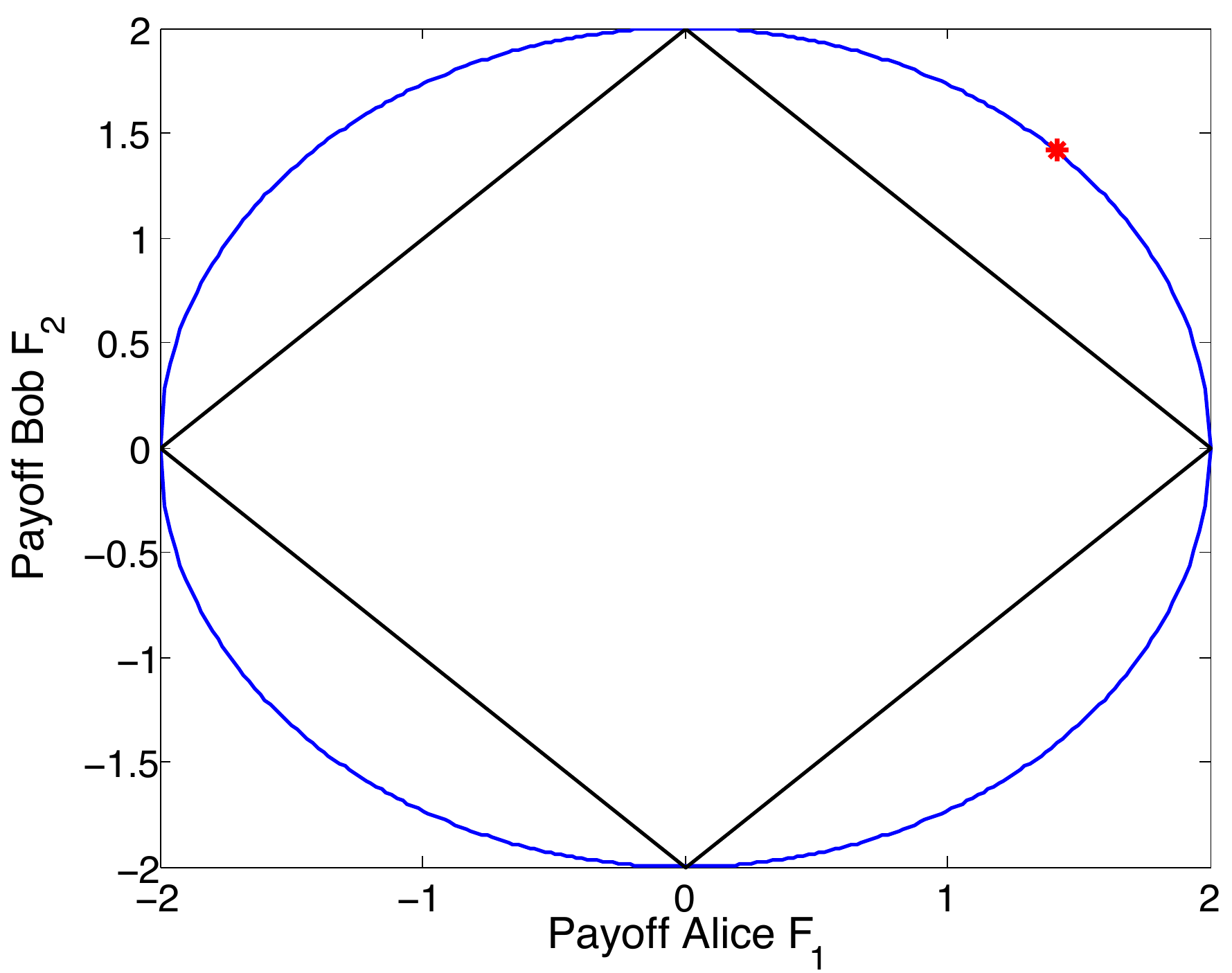}
  \caption{Space of possible average payoffs functions for the Bayesian game of example 2. The pairs of payoffs achievable with a classical advice are represented by the back polytope (square). Players sharing a quantum advice have access to a strictly larger set of possible payoff function, and hence have an advantage over classical players. The red star represents the particular quantum strategy mentioned in the text. Players having access to super-quantum advice can reach all points, in particular the point $F_1=F_2=2$, a no-signaling Nash equilibrium.}
\label{fig}
\end{figure}

{\bf \emph{Example 2.}} Let us now consider an asymmetric variation of the above game. We take again $\mu(X_1,X_2)=1/4$ for $X_1,X_2=0,1$. The payoff functions for Alice and Bob will now be different:
\ba \nonumber f_1(X_1,X_2,A_1,A_2) &=& 
\begin{cases}
4(1-X_1) & \text{if $ A_1 \oplus A_2 = X_1 X_2$} \\
-4(1-X_1) & \text{otherwise}
\end{cases} \\ \nonumber
 f_2(X_1,X_2,A_1,A_2) &=& 
\begin{cases}
+4X_1 & \text{if $ A_1 \oplus A_2 = X_1 X_2$} \\
-4X_1 & \text{otherwise}
\end{cases}
\ea
Hence we obtain the following average payoffs
\ba F_1 = E(X_1=X_2=0) +  E(X_1=0,X_2=1) \\ \nonumber
F_2 = E(X_1=1,X_2=0) -  E(X_1=X_2=1) \ea
It is straightfoward to see that $F_1\leq 2$ and $F_2\leq 2$ for any possible strategy (classical and quantum). However, in the case of classical advice, it holds that $F_1+F_2 \leq 2$, which is simply the CHSH Bell inequality. Note that this inequality is a facet of the space of payoffs. Using quantum advice, in particular the optimal CHSH strategy given in \eqref{Qopt}, one has that $F_1=F_2= \sqrt{2}<2$, but $F_1+F_2=2\sqrt{2}>2$. Thus, we obtain a set of average payoffs which cannot be obtained classically, although each payoff is individually compatible with a classical model (see Fig.~2).

Note that the space of payoff functions in the case of quantum advice is not a polytope in general. Here it can be checked that all points satisfying $F_1^2+F_2^2=4$ can be attained by performing judicious measurements on a singlet state.

Finally, considering advice based on super-quantum correlations leads to even better strategies. Again, the PR box allows both players to reach the optimal payoff, i.e. achieving $F_1=F_2=2$. This point is a no-signaling Nash equilibrium.

{\bf \emph{Example 3.}} Our final example will be more concrete. The players are two companies, both interested in buying jointly some pieces of land, potentially rich in a certain resource. Company A has expertise in extracting this resource, while company B has expertise in selling and distributing it. Hence an association is potentially profitable for both companies, which would then share the net profit equally. 

For company A, the price of production may vary, depending on various parameters, which represent the type of company A. Here the extraction cost can be either low or high (with equal probability), which an expert of company A can evaluate. For company B, the type is the supply on the market, which can be either low or high (with equal probability), evaluated by an expert of company B. Note that the type of each company is private, since each company is reluctant to let the other know about much benefit it could potentially make.

The sale is organized as follows. Both companies will be asked simultaneously, to bid or not on a particular piece of land. If at least one company bids, the piece of land is sold. All money that is bid is retained, hence if both companies bid, their profit is lower than if only one of them bids.
If both the extraction cost and supply on the market are low, the profit will be high. If the extraction cost is low but the supply is high (or conversely) the profit is medium. If both the extraction cost and supply are high, the companies go bankrupt if they bid.
The payoff functions for this Bayesian game are given in Fig.~3. 

It is not difficult to see that, in the case  the companies have access to classical advice, the largest possible average payoff for each company is $F_{A,B}=3/2$. However, having access to quantum advice, the companies can achieve $F_{A,B} \simeq 1.5365>3/2$. This is achieved by performing suitably chosen local measurements on a singlet state. Moreover, this point represents a quantum correlated Nash equilibrium, since no better score can be achieved.

 \begin{figure}[t!]
  \includegraphics[width=\columnwidth]{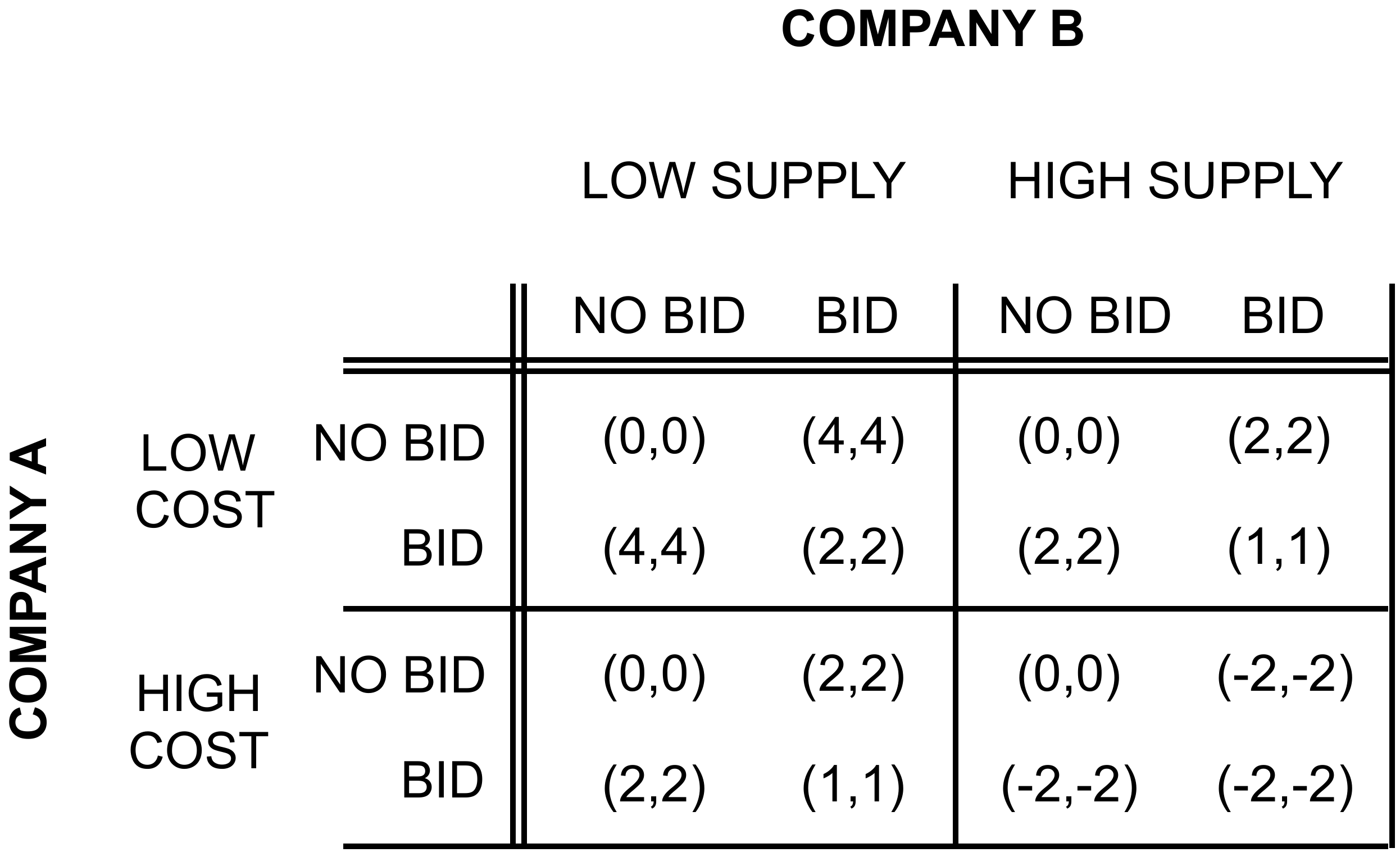}
  \caption{Payoffs for the Bayesian game of example 3.}
\label{fig}
\end{figure}

\section{Discussion}

We have discussed a strong connection between Bell nonlocalty and Bayesian games. This lead us to see that players sharing advice based on nonlocal correlations, for instance using quantum entanglement, can outperform players sharing (any possible) classical advice. Considering the quantum case, it is important to emphasize that the advantage provided by quantum resources is here fully general. Hence it does not rely on any specific restrictions, contrary to previous approaches to quantum games \cite{eisert} which then lead to controversy \cite{comment}. The main point is that these approaches focused on games with complete information (such as Prisoner's dilemma), where the notion of type is not present, in contrast to Bayesian games. 
This is perhaps expressed even more clearly from the point of view of nonlocality: a Bell test can separate quantum from classical predictions only if each observer can choose between several possible measurements to perform.

Finally, we believe that the connection presented here may also benefit nonlocality. Besides providing new potential applications for quantum nonlocality, along with quantum communications \cite{ekert} and communication complexity \cite{cc}, it also raises interesting issues, in particular the possibility of detecting nonlocal correlations via a set of Bell type inequalities, rather than from a single Bell parameter.


\emph{Acknowledgements.} We are very grateful for illuminating discussions with David Leslie and John McNamara. The authors acknowledge financial support from the UK EPSRC, the EU DIQIP, the Swiss National Science Foundation (grant PP00P2\_138917), and the Templeton Foundation.

\end{document}